\documentclass{extarticle}
\usepackage{anyfontsize}
\usepackage{lmodern} 

\usepackage{setspace}
\usepackage{tabularx}

\makeatletter
\renewcommand\normalsize{%
  \@setfontsize\normalsize{12.5pt}{16pt}%
}
\normalsize
\makeatother

\usepackage{amsmath}
\usepackage{graphicx}
\usepackage{natbib}
\usepackage{url} 
\usepackage{amssymb}
\usepackage{enumitem}  
\usepackage[a4paper, top=1.4in, bottom=1.4in, left=1.3in, right=1.3in]{geometry}
\usepackage{indentfirst}

\newcommand{\blind}{0}

\newcommand{\bfepsilon}{\mbox{\boldmath $\epsilon$}}

\newcommand{\bftheta}{\mbox{\boldmath $\theta$}}

	 \newcommand{\bfd}{{\bf d}}	
	 \newcommand{\bff}{{\bf f}}	
	 	
		\newcommand{\bfI}{{\bf I}}

	\newcommand{\bfW}{{\bf W}} 	
\newcommand{\bfy}{{\bf y}}

\newcommand{\bfzero}{{\bf 0}}

\newcommand{\tb}{\textbf}

\newtheorem{result}{Result}

\begin{document}

\def\spacingset#1{\renewcommand{\baselinestretch}%
{#1}\small\normalsize} \spacingset{1}


\if0\blind
{
  \title{\bf Nonlocal Prior Mixture-Based Bayesian Wavelet Regression with Application to Noisy Imaging and Audio Data}
  \author{Nilotpal Sanyal\\[5pt]
    \emph{Department of Mathematical Sciences, University of Texas at El Paso,}\\ \emph{nsanyal@utep.edu}
    }
    \date{}
  \maketitle
} \fi

\if1\blind
{
  \bigskip
  \bigskip
  \bigskip
  \begin{center}
    {\LARGE\bf Title}
\end{center}
  \medskip
} \fi

\thispagestyle{empty}

\bigskip
\hrule 
\begin{center} \tb{Abstract} \end{center}
We propose a novel Bayesian wavelet regression approach using a three-component spike-and-slab prior for wavelet coefficients, combining a point mass at zero, a moment (MOM) prior, and an inverse moment (IMOM) prior. This flexible prior supports small and large coefficients differently, offering advantages for highly dispersed data where wavelet coefficients span multiple scales. The IMOM prior’s heavy tails capture large coefficients, while the MOM prior is better suited for smaller non-zero coefficients. Further, our method introduces innovative hyperparameter specifications for mixture probabilities and scale parameters, including generalized logit, hyperbolic secant, and generalized normal decay for probabilities, and double exponential decay for scaling. Hyperparameters are estimated via an empirical Bayes approach, enabling posterior inference tailored to the data. Extensive simulations demonstrate significant performance gains over two-component wavelet methods. Applications to electroencephalography and noisy audio data illustrate the method’s utility in capturing complex signal characteristics. We implement our method in an R package, NLPwavelet ($\ge$ 1.1).

\noindent%
\vfill

\newpage
\spacingset{1} 

\section{Introduction}
\label{sec:intro}
We focus on spike-and-slab mixture models for wavelet-based Bayesian nonparametric regression. There are several existing approaches that consider for the wavelet coefficients various spike-and-slab mixture models, such as mixtures of two Gaussian distributions with different standard deviations \citep{chipman_et_al_1997}, mixtures of a Gaussian and a point mass at \mbox{zero \citep{clyde_et_al_1998, abramovich_et_al_1998, sanyal_ferreira_2012},} mixtures of a heavy-tailed distribution and a point mass at zero \citep{clyde_george_2000, johnstone_silverman_2005}, mixtures of a logistic distribution and a point mass at zero \citep{sousa_2024}, and mixtures of a nonlocal prior and a point mass at zero \citep{sanyal_ferreira_2017}. Nonlocal priors \citep{johnson_rossell_2010} are a class of priors that assign zero probability density in a neighborhood of the null value (often zero) of the parameter, unlike local priors, which are positive everywhere. In contrast to the other works mentioned above that used traditionally used local priors for the wavelet coefficients, ref. \cite{sanyal_ferreira_2017} pioneered the use of nonlocal priors for wavelet regression. Specifically, ref. \cite{sanyal_ferreira_2017} used two different nonlocal priors, namely the moment prior (MOM) and the inverse moment (IMOM) prior, in their mixture model. In this work, we flexibly extend all the previous approaches by proposing a three-component spike-and-slab mixture model for the wavelet coefficients, where, along with a point mass for the spike part, a mixture of nonlocal priors is used to model the slab component. In addition, we introduce novel hyperparameter specifications that are shown to provide improved estimates in extensive simulation experiments with highly dispersed data.

In the Bayesian paradigm, nonlocal priors have been shown to encourage model selection parsimony and selective shrinkage (unlike local priors) for spurious coefficients \citep{johnson_rossell_2012,rossell_telesca_2017}. They create a gap around zero, yielding harder exclusion and lower bias for sizable effects. In contrast, local shrinkage priors provide continuous shrinkage that is particularly suitable for estimation but less effective for variable selection unless combined with explicit selection rules. Previous work \citep{sanyal_et_al_2019} has observed, in~the context of high-dimensional genomics data, that different nonlocal priors provide better support for large and small regression coefficients. In~wavelet regression, the~wavelet coefficients at multiple resolution levels capture both location and scale characteristics of the underlying function \citep{mallat_2003}. If the underlying function is highly dispersed, its energy is spread across a wide range of location or scale components. For such a function, while most of the wavelet coefficients will likely be small or near zero, non-zero wavelet coefficients will span across multiple scales. In other words, no single scale will dominate the wavelet coefficients, as different scales will capture different portions of the signal’s energy. This will lead to significant coefficients at both coarse and fine scales. We conjecture that, if the underlying function is highly dispersed, a mixture of nonlocal priors will provide better support to the distribution of the wavelet coefficients compared to individual nonlocal priors. Specifically, in this work, we consider for the wavelet coefficients a prior that is a mixture of a point mass at zero, a MOM prior, and an IMOM prior.

Our motivation for combining MOM and IMOM priors stems from the fact that they offer complementary strengths in sparse high-dimensional regression. The MOM prior places more mass around moderate non-zero values, thereby offering greater sensitivity to small but meaningful signals. On the other hand, the IMOM prior has heavier tails, allowing it to support large coefficients better. In the context of wavelet regression of highly dispersed data, a single prior, MOM or IMOM, may over-shrink large coefficients or over-allow small noise fluctuations. A mixture of MOM and IMOM provides adaptive flexibility, simultaneously supporting both ends of the coefficient spectrum. This idea echoes adaptive shrinkage principles in sparse Bayesian learning. From a signal processing perspective, the heterogeneous scaling behavior of wavelet coefficients for highly dispersed signals motivates a prior that can adaptively vary shrinkage across scales. This hybrid slab also enables the prior predictive distribution to cover a broader range of plausible signals, reducing prior-data conflict. While our empirical results support this mixture’s performance, the theoretical justification lies in its capacity to model heterogeneity in signal strengths, a feature not adequately captured by any one nonlocal prior alone.

In Bayesian wavelet regression, the probability weights associated with different component distributions of a spike-and-slab mixture prior, henceforth called \emph{mixture probabilities}, and the scaling parameters of the component distributions are often governed by multiple hyperparameters \citep{abramovich_et_al_1998,clyde_et_al_1998,sanyal_ferreira_2017}. For the mixture probabilities at different resolution levels, previous work considered exponential decay specification \citep{abramovich_et_al_1998}, Bernoulli distribution \citep{clyde_et_al_1998}, and logit specification \citep{sanyal_ferreira_2017}. Further, for the scaling parameters, exponential decay \citep{abramovich_et_al_1998} and polynomial decay specifications \citep{sanyal_ferreira_2017} have been considered. In this work, we propose several novel specifications that flexibly model the variations in the mixture probabilities and scaling components. For the mixture probabilities, we consider a generalized logit decay, a hyperbolic secant decay, and a generalized normal decay, whereas for the scaling parameter, we consider a double exponential decay. Each of these specifications is controlled by a few hyperparameters. Following an empirical Bayes approach, we estimate the hyperparameters from the data and develop a posterior inference conditional on the hyperparameter estimates. 

Through extensive simulation studies, we assess the performance gains of our approach under the different hyperparameter specifications, comparing it to two-component spike-and-slab prior-based wavelet methods. Finally, applications to real-world data, including electroencephalography (EEG) data from a meditation study and audio data from a noisy musical recording, illustrate the practical utility of the proposed method.

Although this work focuses on Bayesian spike-and-slab priors for wavelet analysis, it is still relevant to acknowledge other types of priors explored in the wavelet literature. These include scale mixtures of Gaussians \citep{vidakovic_1998, vidakovic_ruggeri_2001, portilla_et_al_2003, cutillo_et_al_2008}, hidden Markov model-based priors \citep{crouse_et_al_1998}, the generalized Gaussian distribution prior \citep{chang_et_al_2000}, Jeffreys' prior \citep{figueiredo_nowak_2001}, the Bessel K Form prior \citep{boubchir_boashash_2013}, the double Weibull prior \citep{remenyi_vidakovic_2015}, Bayes factor thresholding based on mixtures of conjugate priors \citep{afshari_et_al_2017}, the beta prior \citep{sousa_et_al_2021}, and the logistic prior \citep{sousa_2022}. For comparative reviews and summaries of wavelet-based nonparametric regression methods, see \cite{vidakovic_1999,antoniadis_et_al_2001}.

In what follows, Section~\ref{sec:model} describes the proposed Bayesian hierarchical model along with the hyperparameter specifications, and Section~\ref{sec:inference} describes the empirical Bayes inference procedure. Subsequently, Section~\ref{sec:simulation} describes the simulation experiment and results. The real data applications appear in Section~\ref{sec:EEG} and Section~\ref{sec:audio}. Finally, Section~\ref{sec:discussion} concludes with an overall discussion of our work and relevant remarks. Theoretical proofs are included in Appendix~\ref{app}.


\section{Bayesian Hierarchical~Model}
\label{sec:model}

\subsection{Observation Model}
Suppose $y_1,\ldots,y_n$ represent $n$ noisy observations from an unknown function $f(t)$, which we aim to estimate. We consider the observation model
$
y_i = f(t_i) + \epsilon_i, \quad i=1,\ldots,n,
$
where $t_i=i/n$ represents the equispaced sampling points, and~the errors $\epsilon_i,\ldots,\epsilon_n$ are assumed to be independent and identically distributed (i.i.d.) normal random variables, $\epsilon_i \sim \mathcal{N}(0,\sigma^2)$, with~unknown variance $\sigma^2$. Let $\bfy=(y_1,\ldots,y_n)$ denote the vector of observations, $\bff=(f(t_1),\ldots,f(t_n))$ the vector of true functional values, and~$\bfepsilon=(\epsilon_1,\ldots,\epsilon_n)$ the vector of errors. The observation model can be expressed in matrix form as
\begin{align}
\bfy = \bff + \bfepsilon, \quad \bfepsilon \sim \mathcal{N}(\bfzero,\sigma^2\bfI_n),
\end{align}
where $\bfI_n$ is the $n$-dimensional identity matrix. In~wavelet regression, the~goal is to estimate the function $f(t)$ from the observations $\bfy$ by decomposing $f(t)$ into wavelet basis functions. This decomposition allows us to exploit the multiscale nature of wavelets to capture both local and global features of the function. A~chief feature of wavelet regression is its ability to adapt to different levels of smoothness and to handle noisy data efficiently, making it especially useful in nonparametric function estimation~problems.\\

\subsection{Wavelet Coefficient Model}
We represent $\bff$ using an orthogonal wavelet basis matrix as $\bff=\bfW\bfd$ \citep{donoho_johnstone_1995}, where $\bfW$ is the orthogonal basis matrix and $\bfd$ is a vector whose elements include the scaling coefficient at the coarsest resolution level along with the wavelet coefficients at all resolution levels. Let $\widehat{\bfd}=\bfW^T\bfy$ denote the vector of empirical wavelet coefficients. Since $\bfW$ is orthogonal, we can express $\widehat{\bfd}$ as $\widehat{\bfd}=\bfd + \bfepsilon^*$, where $\bfepsilon^*=\bfW^T\bfepsilon$ represents the transformed error vector with $\bfepsilon^* \sim \mathcal{N}(\bfzero,\sigma^2\bfI_n)$.

Suppose $d_{lj}$ denotes the wavelet coefficient at position $j$ and resolution level $l$, with~$\widehat{d}_{lj}$ defined similarly for the empirical wavelet coefficients. Then, the model in terms of individual coefficients is given by

\begin{align}
\widehat{d}_{lj} = d_{lj} + \epsilon^{*}_{lj}, \quad \epsilon^{*}_{lj} \sim N(0,\sigma^{2})
\end{align}

\subsection{Mixture Prior for the Wavelet Coefficients}
For the wavelet coefficient $d_{lj}$, we consider a spike-and-slab mixture prior that is a mixture of three components---a MOM prior, an~IMOM prior, and~a point mass at zero---given by
\begin{align}
&d_{lj} | \gamma_{l}^{(1)}, \gamma_{l}^{(2)}, \tau_{l}^{(1)}, \tau_{l}^{(2)}, \sigma^{2}, r, \nu \; \sim \; \gamma_{l}^{(1)} MOM\left(\tau_{l}^{(1)},r,\sigma^{2}\right) \notag \\ 
&\qquad\qquad + \left(1 - \gamma_{l}^{(1)}\right) \gamma_{l}^{(2)} IMOM\left(\tau_{l}^{(2)},\nu,\sigma^{2}\right)  + \left(1 - \gamma_{l}^{(1)}\right)\left(1 - \gamma_{l}^{(2)}\right) \delta_{0}(\cdot), \notag \\
&\hspace{9.5cm} 0<\gamma_{l}^{(1)},\gamma_{l}^{(2)}<1,
\end{align}
where $\gamma^{(l)}_1$ and $\gamma^{(2)}_l$ are mixture probabilities. The~$MOM$ prior with order $r$ and variance component $\tau^{(1)}_l\sigma^2$, involving the scale parameter $\tau^{(1)}_l$, has the following density function:
$$
mom(d_{lj} | \tau_{l}^{(1)},r,\sigma^{2}) = \widetilde{M}_{r} \left(\tau_{l}^{(1)}\sigma^{2}\right)^{-r-1/2} d_{lj}^{2r} \exp\left(-\frac{d_{lj}^{2}}{2\tau_{l}^{(1)}\sigma^{2}}\right), \quad r>1, \tau_{l}^{(1)}>0,
$$
where $\widetilde{M}_{r} = (2\pi)^{-1/2}/(2r-1)!!$ and $(2r-1)!!=1 \times 3 \times\ldots \times (2r-1)$. The $IMOM$ prior with shape parameter $\nu$ and variance component $\tau^{(2)}_l\sigma^2$,  involving the scale parameter $\tau^{(2)}_l$, has the following density function:
$$
imom(d_{lj} | \tau_{l}^{(2)},\nu,\sigma^{2}) = \frac{\left(\tau_{l}^{(2)}\sigma^{2}\right)^{\nu/2}}{\Gamma(\nu/2)} |d_{lj}|^{-\nu-1} \exp\left( -\frac{\tau_{l}^{(2)}\sigma^{2}}{d_{lj}^{2}} \right),\quad \nu>1, \tau_{l}^{(2)}>0.
$$

The conditional density of the wavelet coefficient $d_{lj}$ given that $d_{lj}\ne 0$ can be expressed as
\begin{align*}
\pi(d_{lj} | d_{lj}\neq 0) &= \frac{\gamma^{(1)}_l}{\gamma^{(1)}_l + \left(1-\gamma^{(1)}_l\right)\gamma^{(2)}_l} \; \widetilde{M}_{r} \left(\tau_{l}^{(1)}\sigma^{2}\right)^{-r-1/2} d_{lj}^{2r} \exp\left(-\frac{d_{lj}^{2}}{2\tau_{l}^{(1)}\sigma^{2}}\right) \\[5pt]
&\qquad  \frac{\left(1-\gamma^{(1)}_l\right)\gamma^{(2)}_l}{\gamma^{(1)}_l + \left(1-\gamma^{(1)}_l\right)\gamma^{(2)}_l} \; \frac{\left(\tau_{l}^{(2)}\sigma^{2}\right)^{\nu/2}}{\Gamma(\nu/2)} |d_{lj}|^{-\nu-1} \exp\left( -\frac{\tau_{l}^{(2)}\sigma^{2}}{d_{lj}^{2}} \right).
\end{align*}

\begin{figure}[htbp!]
\centering
\includegraphics[width=0.85\textwidth]{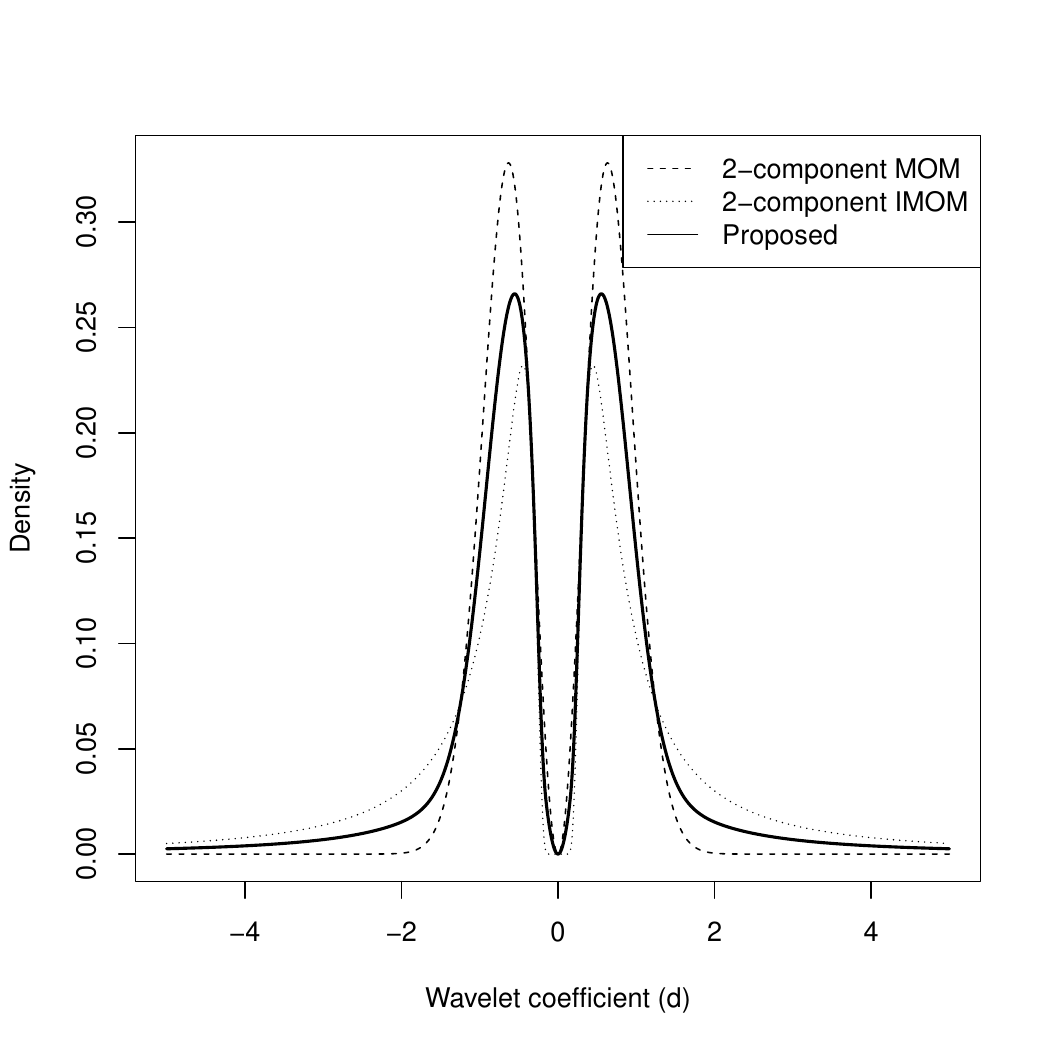}
\caption{Plots of our proposed three-component spike-and-slab mixture model (solid line) with $\gamma_{1l}=\gamma_{2l}=0.25$, $\tau_{1l}=\tau_{2l}=0.2$, $r=\nu=1$, and~$\sigma=1$ along with two-component spike-and-slab mixture models based on MOM (dashed line) and IMOM (dotted line) priors with $\gamma_l=0.5$, $\tau_l=1$, and $\sigma=1$.}
\label{fig:mixture_prior}
\end{figure}

Figure~\ref{fig:mixture_prior} shows plots of our proposed three-component spike-and-slab mixture model (solid line) for $d_{lj}$ with $\gamma_{1l}=\gamma_{2l}=0.25$, $\tau_{1l}=\tau_{2l}=0.2$, $r=\nu=1$, and~$\sigma=1$ along with two-component spike-and-slab mixture models based on MOM (dashed line) and IMOM (dotted line) priors with $\gamma_l=0.5$, $\tau_l=1$, and~$\sigma=1$.

\subsection{Hyperparameter Specifications}
\begin{figure}[htbp!]
\centering
\includegraphics[width=\linewidth]{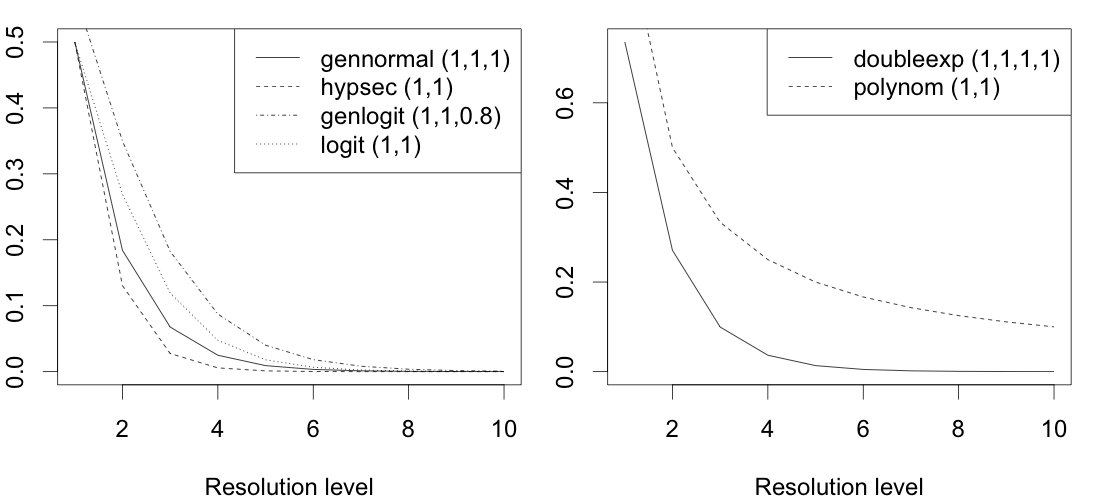}
\caption{Plots of the different specifications considered for the mixture probabilities $\gamma_l^{(1)}$ and $\gamma_l^{(2)}$ (logit, generalized logit, hyperbolic secant, and~generalized normal) and scale parameters $\tau_l^{(1)}$ and $\tau_l^{(2)}$ (polynomial decay and double exponential decay) against resolution level, with specified values of the hyperparameters.}
\label{fig:hyper_specs}
\end{figure}

The mixture prior given in (3) depends on the mixture probabilities $\gamma^{(l)}_1$ and $\gamma^{(2)}_l$, scale parameters $\tau^{(1)}_l$ and $\tau^{(1)}_l$, order $r$, and~shape parameter $\nu$. This section considers different specifications for $\gamma^{(l)}_1$, $\gamma^{(2)}_l$, $\tau^{(1)}_l$, and $\tau^{(1)}_l$. Previous work in wavelet regression considered two-component spike-and-slab mixture priors, where the mixture probabilities were specified using exponential decay specification \citep{abramovich_et_al_1998}, Bernoulli distribution \citep{clyde_et_al_1998}, and~logit specification \citep{sanyal_ferreira_2017}. In this work, we examine three specifications for the mixture probabilities that flexibly model the variations in the mixture probabilities and, to the best of our knowledge, are hitherto unused in wavelet regression. These novel specifications were motivated by an exploratory data analysis where, for multiple highly dispersed data, we observed how the wavelet coefficients changed with resolution level. We compare these novel specifications with  logit specification for $\gamma_{l}^{(1)}$ and $\gamma_{l}^{(2)}$, given by 
$\gamma_l^{(1)} = \exp(\theta^{\gamma}_{1} - \theta^{\gamma}_{2}l) / \{1 + \exp(\theta^{\gamma}_{1} - \theta^{\gamma}_{2}l)\},
\gamma_l^{(2)} = \exp(\theta^{\gamma}_{3} - \theta_{4}l) \ \{1 + \exp(\theta^{\gamma}_{3} - \theta^{\gamma}_{4}l)\},$
where $\theta^{\gamma}_{1},\theta^{\gamma}_{3} \in \mathbb{R}, \; \theta^{\gamma}_{2},\theta^{\gamma}_{4} > 0$ \citep{sanyal_ferreira_2017}. Figure~\ref{fig:hyper_specs} shows, in~the left panel, the~plots of the different specifications for the mixture probabilities against resolution level, with~specific values of the hyperparameters. The novel specifications are described as follows:

\begin{enumerate}[label=(\alph*)]
\item Generalized logit or Richards decay specifications, given by
\begin{align*}
\gamma_l^{(1)} &= \frac{1}{[1 + \exp\{-(\theta^{\gamma}_{1} - \theta^{\gamma}_{2}l)\}]^{\theta^{\gamma}_{3}}}, \quad 
\theta^{\gamma}_{1} \in \mathbb{R}, \; \theta^{\gamma}_{2},\theta^{\gamma}_{3} > 0 \\[5pt]
\gamma_l^{(2)} &= \frac{1}{[1 + \exp\{-(\theta^{\gamma}_{4} - \theta^{\gamma}_{5}l)\}]^{\theta^{\gamma}_{6}}}, \quad
\theta^{\gamma}_{4} \in \mathbb{R}, \; \theta^{\gamma}_{5},\theta^{\gamma}_{6} > 0.
\end{align*}
This form corresponds to a flexible S-shaped decay curve that reduces to standard logistic decay when $\theta^{\gamma}_{3}$ (or $\theta^{\gamma}_{6}$) and controls the steepness of the curve equal to one.

\item Hyperbolic secant decay specifications, given by
\begin{align*}
\gamma_l^{(1)} &= \frac{2}{\pi} \arctan\left[\exp\left(\frac{\pi}{2} \left(\theta^{\gamma}_{1} - \theta^{\gamma}_{2}l\right)\right)\right], \quad
\theta^{\gamma}_{1} \in \mathbb{R}, \; \theta^{\gamma}_{2} > 0 \\[5pt]
\gamma_l^{(2)} &= \frac{2}{\pi} \arctan\left[\exp\left(\frac{\pi}{2} \left(\theta^{\gamma}_{3} - \theta^{\gamma}_{4}l\right)\right)\right], \quad
\theta^{\gamma}_{3} \in \mathbb{R}, \; \theta^{\gamma}_{4} > 0.
\end{align*}
This form, although less intuitive, is also sigmoid-like but with heavier tails and slower decay than logistic, indicating that the mixture parameters approach one more slowly with a smoother transition. $\theta^{\gamma}_{2}$ (or $\theta^{\gamma}_{4}$) controls steepness and $\theta^{\gamma}_{1}$ (or $\theta^{\gamma}_{3}$) controls shift.

\item Generalized normal decay specifications, given by
\begin{align*}
\gamma_l^{(1)} &= \frac{1}{2} + \text{sign}(\theta^{\gamma}_{1}-l) \frac{1}{2\Gamma(1/\theta^{\gamma}_{2})} \; \gamma\left(1/\theta^{\gamma}_{2} ,\left|\frac{\theta^{\gamma}_{1}-l}{\theta^{\gamma}_{3}}\right|^{\theta^{\gamma}_{2}}\right), \quad 
\theta^{\gamma}_{1} \in \mathbb{R}, \; \theta^{\gamma}_{2},\theta^{\gamma}_{3} > 0 \\[5pt]
\gamma_l^{(2)} &= \frac{1}{2} + \text{sign}(\theta^{\gamma}_{4}-l) \frac{1}{2\Gamma(1/\theta^{\gamma}_{5})} \; \gamma\left(1/\theta^{\gamma}_{5} ,\left|\frac{\theta^{\gamma}_{4}-l}{\theta^{\gamma}_{6}}\right|^{\theta^{\gamma}_{5}}\right), \quad
\theta^{\gamma}_{4} \in \mathbb{R}, \; \theta^{\gamma}_{5},\theta^{\gamma}_{6} > 0.
\end{align*}
This form is most flexible and can mimic normal CDF, Laplace CDF, and many other distributions, but also most complex. Whereas $\theta^{\gamma}_{2}$ (or $\theta^{\gamma}_{2}$) controls tail behavior with larger values indicating lighter tails, others control scale and shift.
\end{enumerate}

For scale parameters of spike-and-slab mixture priors, previously considered specifications include exponential decay \citep{abramovich_et_al_1998} and polynomial decay \citep{sanyal_ferreira_2017}. Here, for~$\tau_{l}^{(1)}$ and $\tau_{l}^{(2)}$, we consider the polynomial decay specification given by 
$\tau_{l}^{(1)} = \theta^{\tau}_{1}l^{-\theta^{\tau}_{2}},  
\tau_{l}^{(2)} = \theta^{\tau}_{3}l^{-\theta^{\tau}_{4}}$
with $\theta^{\tau}_{1},\theta^{\tau}_{2},\theta^{\tau}_{3},\theta^{\tau}_{4} > 0$. In addition, for modeling the variations in the scale parameters more flexibly, we novelly propose the following:  
\begin{enumerate}[label=(\alph*)]
\item[(d)] Double exponential decay specifications, given by
\begin{align*}
\tau_{l}^{(1)} = \theta^{\tau}_{1} \exp(-\theta^{\tau}_{2} l) + \theta^{\tau}_{3} \exp(-\theta^{\tau}_{4} l), \quad
\theta^{\tau}_{1},\theta^{\tau}_{2},\theta^{\tau}_{3},\theta^{\tau}_{4} > 0 \\[5pt] 
\tau_{l}^{(2)} = \theta^{\tau}_{5} \exp(-\theta^{\tau}_{6} l) + \theta^{\tau}_{7} \exp(-\theta^{\tau}_{8} l), \quad
\theta^{\tau}_{5},\theta^{\tau}_{6},\theta^{\tau}_{7},\theta^{\tau}_{8} > 0
\end{align*}
\end{enumerate}

Figure~\ref{fig:hyper_specs} shows, in the right panel, the plots of the different specifications for the scale parameters against resolution level, with specific values of the hyperparameters.

\unskip
\vspace{-6pt}

Each of the proposed specifications admits interpretable controls over rate and shape of decay across scales, crucial for adaptivity in multi-resolution modeling, and is governed by a small number of hyperparameters. Note that the hyperparameters for the mixture probabilities are superscripted with $\gamma$ and those for the scale parameters are superscripted with $\tau$. In our simulation studies described in Section~\ref{sec:simulation}, we analyze each simulated dataset using $4\times 2=8$ configurations arising out of the combinations of the above specifications for the mixture probabilities and the scale parameters. While no formal optimality results are available for these specific forms, their empirical performance, as shown in Section~\ref{sec:simulation}, supports their practical relevance. Let $\bftheta$ generically denote the set of all hyperparameters for any given configuration.

\section{Inference} 
\label{sec:inference}
For inference in our proposed Bayesian hierarchical wavelet regression model based on three-component spike-and-slab mixture priors, we adopt the empirical Bayes approach \citep{clyde_george_2000, sanyal_ferreira_2017}. This methodology estimates the hyperparameters from the data and performs posterior inference conditioned on these estimated hyperparameters. For simplicity, in our simulation analysis, we set $r=1$ and $\nu=1$. Further, we estimate the error variance $\sigma^2$ using the median absolute deviation estimator \citep{donoho_et_al_1995} $\hat{\sigma}={0.6745}^{-1}\; \text{median}_j(|\widehat{d}_{Lj} - \text{median}_j(\widehat{d}_{Lj})|)$, which is a well-established practice in wavelet regression \citep{abramovich_et_al_1998, clyde_et_al_1998, clyde_george_2000, johnstone_silverman_2005}.

\subsection{Hyperparameter Estimation}
Let $\delta(x)$ denote the value of the point mass function $\delta(0)$ at $x$. The following result is used to obtain the hyperparameter estimates.

\begin{result}
Integrating out $d_{lj}$ from the wavelet coefficient model (2) using the mixture prior of the wavelet coefficients in (3) and using the Laplace approximation for the IMOM prior component, we get the marginal distribution of the empirical wavelet coefficients, $\widehat{d}_{lj}$, as
\begin{align}
&\pi(\widehat{d}_{lj} | \sigma^{2},\bftheta,r,\nu) \approx \gamma_{l}^{(1)} \left(1+\tau_{l}^{(1)}\right)^{-r} M_{r}^{*}(\widehat{d}_{lj},\tau_{l}^{(1)},\sigma^{2}) \; \phi\left(\widehat{d}_{lj};0,\sigma^{2}\left(1+\tau_{l}^{(1)}\right)\right) \notag\\
&\hspace{3.5cm}+ \left(1 - \gamma_{l}^{(1)}\right)\gamma_{l}^{(2)}  \frac{\left(\tau_{l}^{(2)}\sigma^{2}\right)^{\nu \over 2}}{\Gamma(\nu/2)} \phi\left(\widehat{d}_{lj};0,\sigma^{2}\right)  \sqrt{2\pi} \sigma_{*} h(d_{lj}^{*}(\widehat{d}_{lj})) \notag\\[5pt]
&\hspace{3.55cm} + \left(1 - \gamma_{l}^{(1)}\right)\left(1 - \gamma_{l}^{(2)}\right) \phi\left(\widehat{d}_{lj};0,\sigma^{2}\right),
\end{align}
where, in the offshoot of the MOM component,
$$
M_r^{\star} \left(\widehat{d}_{lj}, \tau_l^{(1)}, \sigma^2\right) = \frac{1}{(2r-1)!!} \sum_{i=0}^r \frac{(2r)!}{(2i)!(r-i)!2^{r-i}} \left( \sqrt{\frac{\tau^{(1)}_l}{1+\tau^{(1)}_l}}\frac{\widehat{d}_{lj}}{\sigma} \right)^{2i},
$$
and in the offshoot of the IMOM component,
$$
h(d_{lj}) = |d_{lj}|^{-(\nu+1)} \exp\left\{ -\frac{1}{2\sigma^2} \left(d_{lj}^2 - 2d_{lj}\widehat{d}_{lj}\right) - \frac{\tau_l^{(2)}\sigma^2}{d_{lj}^2} \right\},
$$
$d_{lj}^{*}(\widehat{d}_{lj})$ is the global maxima of $h(d_{lj})$, and~$\sigma_*^2=-1/L_h''(d_{lj}^{*}(\widehat{d}_{lj}))$, with~$L_h(d_{lj})=\log(h(d_{lj}))$.
\end{result}
The proof of Result 1 is given in Appendix \ref{app}. Using (4), the marginal likelihood function is approximately given by

\begin{align*}
&\prod_l\prod_j \left[ \gamma_{l}^{(1)} \left(1+\tau_{l}^{(1)}\right)^{-r} M_{r}^{*}(\widehat{d}_{lj},\tau_{l}^{(1)},\sigma^{2}) \; \phi\left(\widehat{d}_{lj};0,\sigma^{2}\left(1+\tau_{l}^{(1)}\right)\right) \right.\notag\\
&\hspace{2cm}\left.+ \left(1 - \gamma_{l}^{(1)}\right)\gamma_{l}^{(2)}  \frac{\left(\tau_{l}^{(2)}\sigma^{2}\right)^{\nu \over 2}}{\Gamma(\nu/2)} \phi\left(\widehat{d}_{lj};0,\sigma^{2}\right)  \sqrt{2\pi} \sigma_{*} h(d_{lj}^{*}(\widehat{d}_{lj})) \right.\notag\\[5pt]
&\hspace{2.05cm} \left.+ \left(1 - \gamma_{l}^{(1)}\right)\left(1 - \gamma_{l}^{(2)}\right) \phi\left(\widehat{d}_{lj};0,\sigma^{2}\right) \right].
\end{align*}
This is a function only of the data and the hyperparameters. Hence, we maximize this function with respect to the hyperparameters to obtain their estimates, $\widehat{\bftheta}$. With $\bftheta=\widehat{\bftheta}$, the prior distribution in (3) is fully known.

\subsection{Posterior Distribution}
The following two results provide the posterior estimates of the wavelet coefficients.

\begin{result}
The conditional posterior density of the wavelet coefficient $d_{lj}$, given that $d_{lj}\ne 0$ and the hyperparameter estimates $\hat{\bftheta}$, by using the Laplace approximation for the IMOM prior component, can be expressed as
\begin{align*}
&\pi(d_{lj} | d_{lj}\neq 0, \sigma^{2},\bftheta,r,\nu,\bfy) = \frac{p_{lj}^{(1)}}{p_{lj}^{(1)} + p_{lj}^{(2)}} \; \frac{\widetilde{M}_r}{M_r^*\left(\widehat{d}_{lj}, \tau_l^{(1)}, \sigma^2\right)}  \; d_{lj}^{2r}  \\ 
&\qquad\exp\left\{ -\frac{1}{2\sigma^{2}\frac{\tau_{l}^{(1)}}{\left(1+\tau_{l}^{(1)}\right)}} \left( d_{lj} - \frac{\tau_{l}^{(1)}}{1+\tau_{l}^{(1)}} \widehat{d}_{lj} \right)^{2}\right\} + \frac{p_{lj}^{(2)}}{p_{lj}^{(1)} + p_{lj}^{(2)}} \; \phi\left(d_{lj} | d_{lj}^{*}(\widehat{d}_{lj}), \sigma_{*}^{2}\right),
\end{align*}
where
$$
p_{lj}^{(1)} = \frac{O_{lj}^{(1)}}{1+O_{lj}^{(1)}+O_{lj}^{(2)}}, \qquad p_{lj}^{(2)} = \frac{O_{lj}^{(2)}}{1+O_{lj}^{(1)}+O_{lj}^{(2)}},
$$
$$
O_{lj}^{(1)} = \frac{\gamma_{l}^{(1)}}{\left(1-\gamma_{l}^{(1)}\right)\left(1-\gamma_{l}^{(2)}\right)} \left(1+\tau_{l}^{(1)}\right)^{-r-1/2} M_{r}^{*} \exp\left\{ \frac{1}{2\sigma^{2}}  \frac{\tau_{l}^{(1)}}{1+\tau_{l}^{(1)}} \widehat{d}_{lj}^{2} \right\},
$$
and
$$
O_{lj}^{(2)} = \frac{\gamma_{l}^{(2)}}{1-\gamma_{l}^{(2)}} \frac{\left(\tau_{l}^{(2)}\sigma^{2}\right)^{\nu \over 2}}{\Gamma(\nu/2)} \sqrt{2\pi} \sigma_{*} h(d_{lj}^{*}(\widehat{d}_{lj})).
$$
\end{result}
The proof is given in Appendix \ref{app}.

\begin{result}
The posterior expectation of the wavelet coefficients $d_{lj}$ is 
\begin{align*}
\bar{d}_{lj} &= E(d_{lj}|y) = p_{lj}^{(1)}  \frac{M_r^{**}}{M_r^*}  \sqrt{\frac{\tau_l^{(1)}}{1+\tau_l^{(1)}}}  \sigma  +   p_{lj}^{(2)}   d_{lj}^{*}(\widehat{d}_{lj}),
\end{align*}
where
$$
M_r^{\star\star}\left(\widehat{d}_{lj}, \tau_l, \sigma^2\right) =  \frac{1}{(2r-1)!!}  \sum_{i=1}^{r+1} \frac{(2r+1)!}{(2i-1)!(r+1-i)!2^{r+1-i}} \left( \sqrt{\frac{\tau_l^{(1)}}{1+\tau_l^{(1)}}} \frac{\widehat{d}_{lj}}{\sigma} \right)^{2i-1}.
$$
\end{result}
The proof of Result 3 is immediate from the proof of Result 2 and hence is omitted. Let $\bar{\bfd}=\{d_{lj}:l=1,\ldots,L,j=1\ldots,J\}$ denote the vector of posterior means of the wavelet coefficients. We use $\bar{\bfd}$ in the inverse discrete wavelet transform to obtain the posterior mean of the unknown function $f$, given by
$$
\bar{\bff} = \operatorname{E}(\bff | \bfy) = \bfW\operatorname{E}(\bfd | \bfy) = \bfW\bar{\bfd}.
$$

\section{Simulation study} \label{sec:simulation}
In this section, we conduct an extensive simulation analysis to comparatively evaluate the performance of the proposed method with different hyperparameter configurations. We consider three well-known test functions proposed by~\cite{donoho_johnstone_1994}---\emph{blocks}, \emph{bumps}, and \emph{doppler}---that are used as standard test functions in the wavelet literature. However, we modify the coefficients used by~\cite{donoho_johnstone_1994} to obtain more highly dispersed signals. We define our modified test functions as

$$
f_{blocks}(t) = \sum h_j K(t-t_j), \text{ where } K(t)=\{1+\text{sign}(t)\}/2,
$$
$(tj) = (0.1, 0.13, 0.15, 0.23, 0.25, 0.40, 0.44, 0.65, 0.76, 0.78, 0.81)$,\\ 
$(hj) = (4, -8, 3, -4, 8, -4.2, 2.1, 4.3, -6.1, 2.1, -4.7)$;
$$
f_{bumps}(t) = \sum h_j K((t - t_j)/w_j), \text{ where } K(t) = (1 + |t|)^{-4},
$$
$(t_j) = (0.1, 0.13, 0.15, 0.23, 0.25, 0.40, 0.44, 0.65, 0.76, 0.78, 0.81)$,\\ 
$(h_j) = (2, 10, 1, 4, 8, 4.2, 2.1, 4.3, 1.1, 3.1, 8.2)$,\\  
$(w_j) = (0.005, 0.005, 0.006, 0.01, 0.01, 0.03, 0.01, 0.01, 0.005, 0.008, 0.005)$;
$$
f_{doppler}(t) = \{t(1-t)\}^{1/2} \sin\{2\pi(1 + \epsilon)/(t + \epsilon)\}, \epsilon = 0.01.
$$
As these modifications involved adjustments to specific numerical values only, the overall functional forms remained consistent with the original Donoho--Johnstone test functions. In addition, we consider three different linear combinations of these functions---\emph{lcomb1}, \emph{lcomb2}, and~\emph{lcomb3}---that represent curves that combine various features---such as blockiness, bumpiness, and changing frequency---in different proportions, given by
\begin{align*}
f_{lcomb1}(t) &= 0.4 f_{blocks}(t) + 0.4 f_{bumps}(t) + 0.2 f_{doppler}(t) \\
f_{lcomb2}(t) &= 0.4 f_{blocks}(t) + 0.2 f_{bumps}(t) + 0.4 f_{doppler}(t) \\
f_{lcomb3}(t) &= 0.2 f_{blocks}(t) + 0.4 f_{bumps}(t) + 0.4 f_{doppler}(t).
\end{align*}

\begin{figure}[htbp!]
\centering
\includegraphics[width=\linewidth]{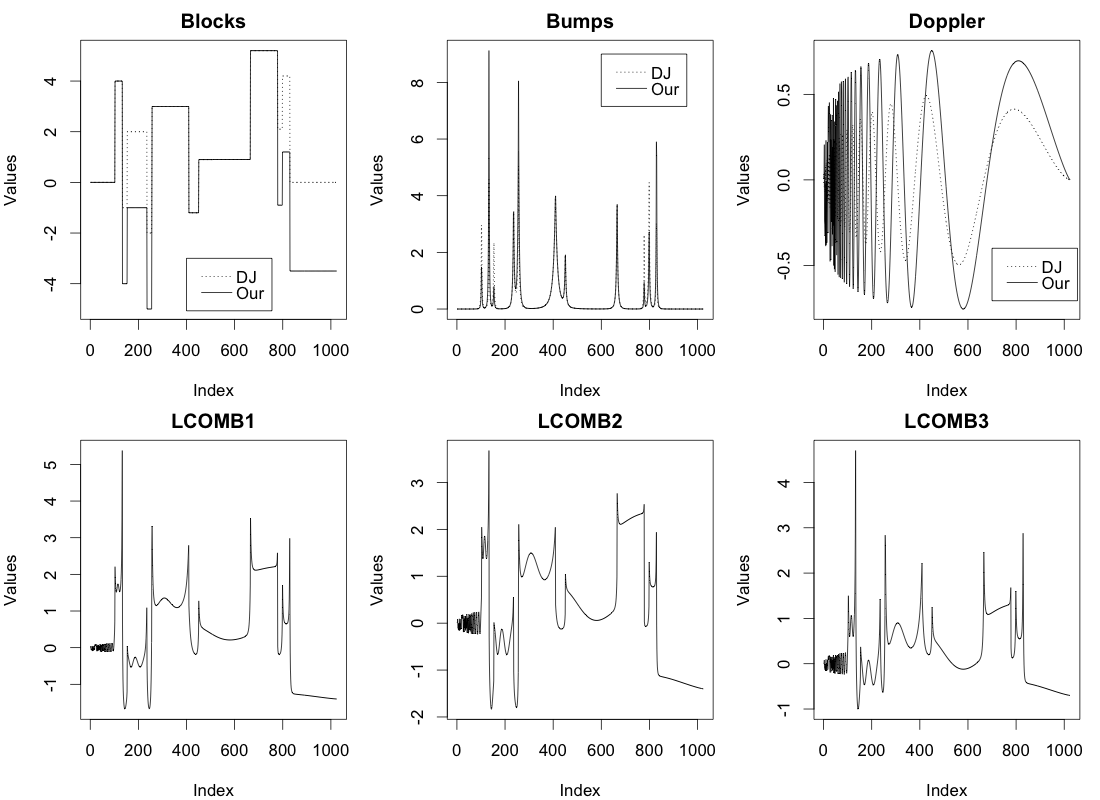}
\caption{Plots of the test functions \emph{blocks}, \emph{bumps}, and \emph{doppler} using the Donoho--Johnstone (DJ) specifications (dashed line) and our specifications (solid line), and three linear combinations of them---\emph{lcomb1}, \emph{lcomb2}, and \emph{lcomb3}---based on our specifications, all evaluated at 1024 equally spaced in (0,1).}
\label{fig:testfunc}
\end{figure}

Figure~\ref{fig:testfunc} shows the plots of the test functions blocks, bumps, and~doppler using Donoho--Johnstone specifications (dashed line) and our specifications (solid line), and \emph{lcomb1}, \emph{lcomb2}, and \emph{lcomb3} based on our specifications, all evaluated at 1024 equally spaced in (0,1). We evaluate each considered test function at $n$ = 512, 1024, 2048, \text{and } 4096~equidistant points in the interval (0,1) and add random Gaussian noise with mean 0 to generate data with signal-to-noise ratio, SNR $= 3, 5, \text{and } 7$. For each combination of $n$ and SNR, we consider 100 replications. The simulated datasets are analyzed using the proposed methodology with eight different hyperparameter configurations described in Section~\ref{sec:model}. For comparison, we also analyzed the datasets using individual MOM and IMOM prior-based two-component mixture models \citep{sanyal_ferreira_2017}. Thus, a total of 24 analysis methods were applied to each dataset. Note that the prior literature \citep{sanyal_ferreira_2017} has already shown that nonlocal prior-based wavelet analysis generally performs better than other existing wavelet-based methods such as sure \citep{donoho_johnstone_1995}, BayesThresh \citep{abramovich_et_al_1998}, cv \citep{nason_1996}, fdr \citep{abramovich_benjamini_1996}, and~Ebayesthresh \citep{johnstone_silverman_2005}. So, for brevity, we do not compare with these methods in this work. For wavelet transformation, we consider the Daubechies least asymmetric wavelet with six vanishing moments and periodic boundary conditions. Wavelet computations are implemented using the R package \emph{wavethresh} \citep{nason_2024}. 

\vspace{-6pt}

Table~\ref{tab:method_compare} presents, for each test function and analysis method, the number of ($n$, SNR) combinations where the method achieved the lowest mean squared error (MSE). For each test function, the method with the highest frequency of best performance (i.e., the lowest MSE) is highlighted in bold. We observe the following:
\begin{enumerate}[label=(\alph*)]
\item For every function, the highest frequency of best performance was shown by a three-component mixture method (with one tie with a two-component mixture method for the blocks function), which is proposed in the current work.
\item Out of all eight three-component mixture methods, the one with generalized normal specification for the mixture probabilities and double exponential decay specification for the scale parameters (\emph{mixture-gennormal-doubleexp}) showed the maximum number of best performances in total (12 times) for all the test functions. This was followed by the method using logit specification for the mixture probabilities and polynomial decay specification for the scale parameters (\emph{mixture-logit-polynom}) (11 times) and the method using generalized normal specification for the mixture probabilities and polynomial decay specification for the scale parameters (\emph{mixture-logit-polynom}) (9 times).
\item Considering only the test functions \emph{lcomb1}, \emph{lcomb2}, and \emph{lcomb3} that represent signals with mixed characteristics in various proportions, the \emph{mixture-gennormal-doubleexp} method showed the maximum number of best performances (10 times).
\end{enumerate}

\begin{figure}[htbp!]
\centering
\includegraphics[width=\textwidth]{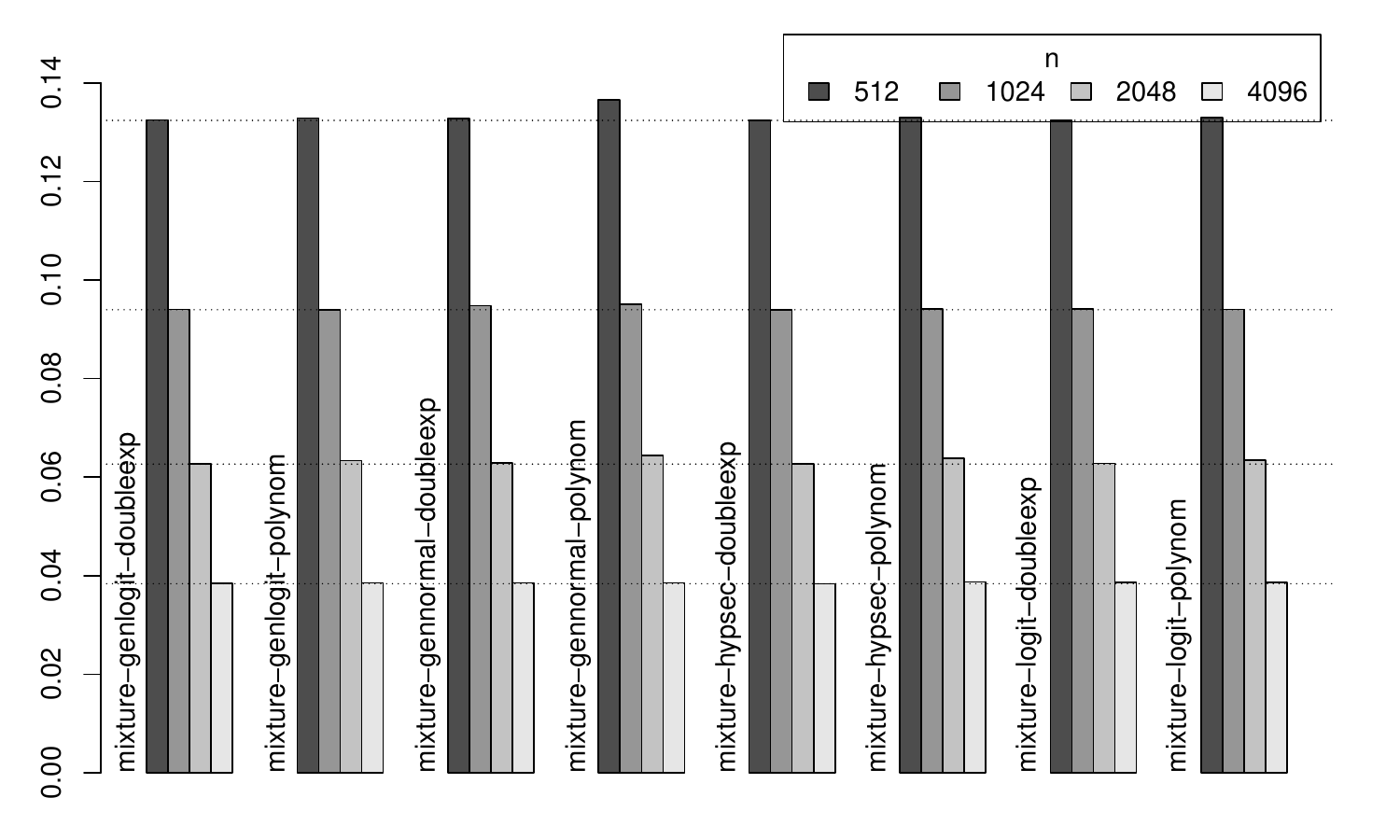}
\vspace{-20pt}
\caption{For the proposed three-component mixture-based methods, MSE for different sample sizes ($n$), averaged over 3 SNRs and 6 test functions.}
\label{fig:MSEplot}
\end{figure}

For real data, SNR and the original function will not be known. So, next, in Figure \ref{fig:MSEplot}, we show, for each of the eight three-component mixture methods, the MSE for the four different sample sizes (n), averaged over the three SNRs and six test functions considered in our study. Overall, methods with a double exponential decay specification for the scale parameters showed a better performance. Specifically, for the two largest sample sizes (2048 and 4096), the method with hyperbolic secant and double exponential decay specifications (\emph{mixture-hypsec-doubleexp}) and the method with generalized logit and double exponential decay specifications (\emph{mixture-genlogit-doubleexp}) provided top performances. Notably, the method with generalized normal and polynomial decay specifications showed a worse performance in Figure~\ref{fig:MSEplot}, even though the methods with generalized normal specifications showed the best performance in Table~\ref {tab:method_compare}. This implies that the methods with generalized normal specifications, although most often provide the best denoising, may sometimes produce large bias that negatively affects their overall average performance. So, the results obtained by their usage should be validated by field knowledge or external means (such as auditory assessment for sound signals).

\begin{table}[htbp!]
\caption{Method comparison: Data were simulated for each test function across 12 combinations of sample size ($n$) and SNR, with 100 replications per combination. A total of 24 analysis methods were applied to each dataset---8 methods with MOM-based two-component mixture prior, 8 methods with IMOM-based two-component mixture prior, and 8 methods with the proposed three-component mixture prior. The average MSE was computed for each method across the replications. The table presents, for each test function and analysis method, the number of $(n, \text{SNR})$ combinations where the method achieved the lowest MSE. For each test function, the method with the highest frequency of best performance is highlighted in bold.\\[5pt] }
\scalebox{0.89}{
\setlength{\tabcolsep}{3pt}
\begin{tabular}{l@{\hskip 0pt}rrrrrrr}
Method & \emph{blocks} & \emph{bumps} & \emph{doppler} & \emph{lcomb1} & \emph{lcomb2} & \emph{lcomb3} & Total \\
\hline
mom-logit-polynom           &  1 & 0 & 0 & 0 & 0 & 0 & 1 \\
mom-logit-doubleexp         &  1 & 0 & 0 & 0 & 0 & 0 & 1 \\
mom-genlogit-polynom        &  0 & 0 & 0 & 0 & 0 & 0 & 0 \\
mom-genlogit-doubleexp      &  0 & 0 & 0 & 0 & 0 & 0 & 0 \\
mom-hypsec-polynom          &  0 & 0 & 0 & 0 & 1 & 0 & 1 \\
mom-hypsec-doubleexp        &  {\tb{3}} & 0 & 0 & 0 & 0 & 0 & 3 \\
mom-gennormal-polynom       &  0 & 0 & 0 & 0 & 0 & 0 & 0 \\
mom-gennormal-doubleexp     &  0 & 0 & 0 & 0 & 0 & 0 & 0 \\[6pt]
imom-logit-polynom          &  0 & 0 & 0 & 0 & 0 & 0 & 0 \\
imom-logit-doubleexp        &  0 & 0 & 0 & 0 & 0 & 0 & 0 \\
imom-genlogit-polynom       &  0 & 0 & 0 & 0 & 0 & 0 & 0 \\
imom-genlogit-doubleexp     &  0 & 0 & 0 & 0 & 0 & 0 & 0 \\
imom-hypsec-polynom         &  0 & 0 & 0 & 0 & 0 & 0 & 0 \\
imom-hypsec-doubleexp       &  0 & 0 & 0 & 0 & 0 & 0 & 0 \\
imom-gennormal-polynom      &  0 & 0 & 0 & 0 & 0 & 0 & 0 \\
imom-gennormal-doubleexp    &  0 & 0 & 0 & 0 & 0 & 0 & 0 \\[6pt]
mixture-logit-polynom       &  1 & {\tb{6}} & 0 & 1 & {\tb{3}} & 0 & 11 \\
mixture-logit-doubleexp     &  0 & 1 & 2 & 1 & 2 & 1 & 7 \\
mixture-genlogit-polynom    &  2 & 1 & 2 & 0 & 1 & 2 & 8 \\
mixture-genlogit-doubleexp  &  {\tb{3}} & 0 & 0 & 0 & 1 & 0 & 4 \\
mixture-hypsec-polynom      &  1 & 0 & {\tb{4}} & 1 & 1 & 0 & 7 \\
mixture-hypsec-doubleexp    &  0 & 1 & 0 & {\tb{4}} & 1 & 2 & 8 \\
mixture-gennormal-polynom   &  0 & 3 & 2 & 1 & 1 & 2 & 9 \\
mixture-gennormal-doubleexp &  0 & 0 & 2 & 4 & 1 & {\tb{5}} & 12 \\ \hline
Total & 12 & 12 & 12 & 12 & 12 & 12 & 72 \\
\end{tabular}
}
\label{tab:method_compare}
\end{table}
\normalsize

\section{EEG Meditation~Study} \label{sec:EEG}
We demonstrate the utility and flexibility of the proposed methodology through the analysis of EEG data from a meditation study, investigating the relation between mind wandering and meditation practice. Detailed information about the study is provided in~\cite{brandmeyer_delorme_2018}, and~the dataset is available on the \emph{OpenNeuro} platform (see the Supplementary Materials Section S1 for the link). The meditation experiment involved 24 subjects---12 experienced meditators (10 males, 2 females) and 12 novices (2 males, 10 females). Participants meditated while being interrupted approximately every two minutes to report their level of concentration and mind wandering via three probing questions. Each participant completed two to three sessions lasting 45 to 90 minutes, with a minimum of 30 probes per participant. EEG data were collected using a 64-channel Biosemi system (channels A1–A32 and B1–B32) with a Biosemi 10--20 head cap montage at a sampling rate of 2048 Hz, providing spatial information about brain activity across different scalp regions. The dataset available on \emph{OpenNeuro} has already been downsampled to 256~Hz.

For our analysis, we focused on the data from session 1 of four participants: two expert meditators (one male and one female, identified as subjects 10 and 5 in the original dataset, respectively) and two novices (one male and one female, identified as subjects 23 and 14, respectively). The data were average-referenced to mitigate common noise and artifacts by calculating the mean signal across all electrodes at each time point and subtracting it from the signal at each electrode. Following this, a~2 Hz high-pass filter was applied using an infinite impulse response (IIR) filter with a transition bandwidth of 0.7 Hz and an order of six. Denoting the time of probe 1 by $t$, we analyzed the EEG signal within the 16 s interval $(t-8,t+8)$. With a sampling rate of 256 Hz, this interval comprised 256$\times$16 = 4096 data points, representing noisy observations of the underlying signal. We analyzed the EEG data using the \emph{mixture-hypsec-doubleexp}, \emph{mixture-genlogit-doubleexp}, and~\emph{mixture-gennormal-polynom} methods (see Section~\ref{sec:simulation}), employing the Daubechies least asymmetric wavelet transform with six vanishing moments and periodic boundary~conditions.

\begin{figure}[htbp!]
\centering
\includegraphics[width=\linewidth]{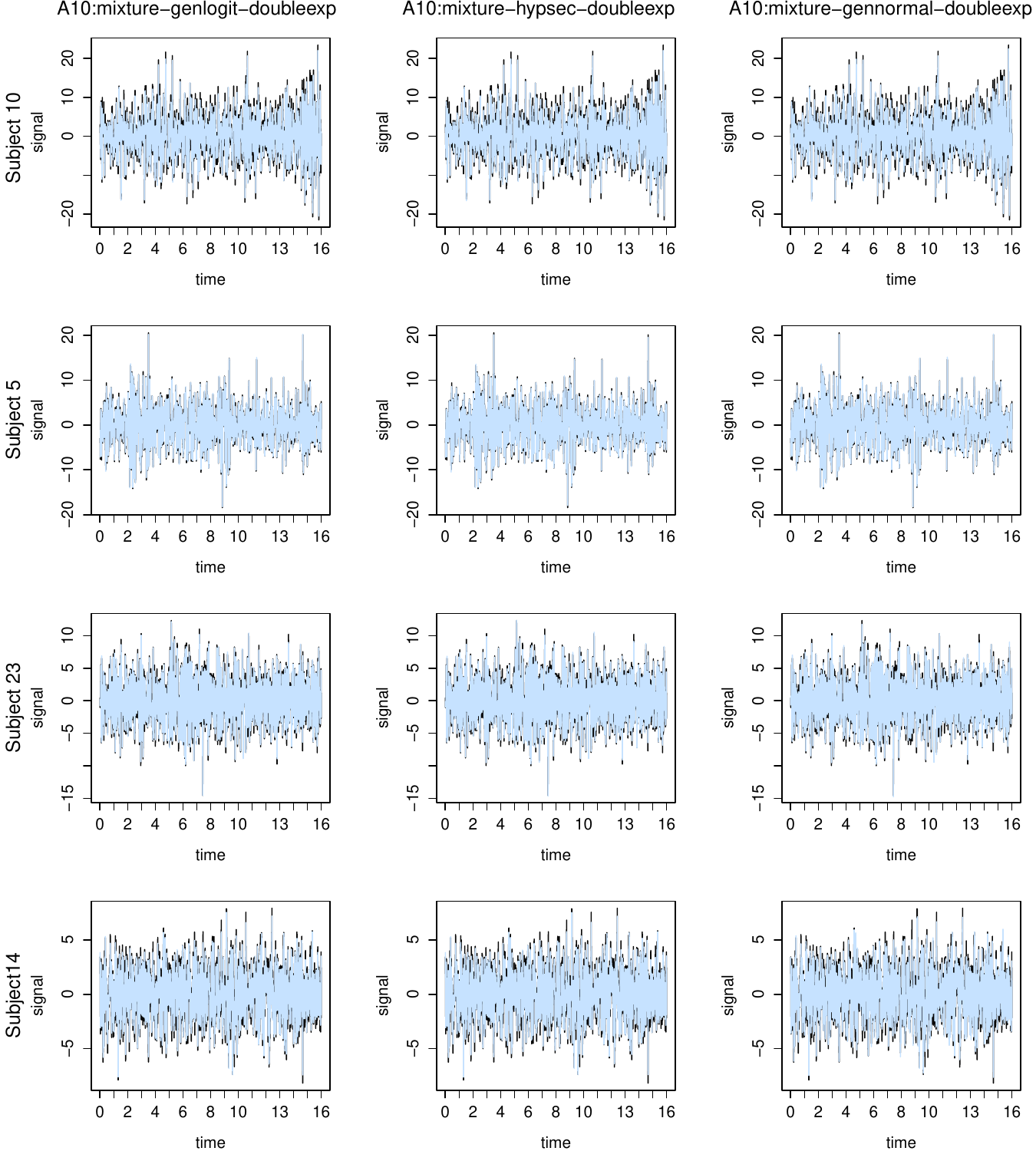}
\caption{Plots of the posterior means of the EEG signal (in slate gray) based on the imom-logit-polynom (\textbf{left}), mixture-genlogit-doubleexp (\textbf{middle}), and~mixture-gennorm-polynom (\textbf{right}) methods, superimposed on the observed data (in black), obtained during the 16 s interval $(t-8,t+8)$, $t$ being probe 1 onset time, from the A10 channel of the 4 considered participants.}
\label{fig:EEG_post}
\end{figure}

Figure~\ref{fig:EEG_post} presents the plots of the posterior means of the EEG signal (in slategray) based on the mixture-genlogit-doubleexp (left column), mixture-hypsec-doubleexp (middle column), and mixture-gennormal-doubleexp (right column) methods, superimposed on the observed data (in black), obtained during the 16 s interval from the A10 channel of the four considered participants. Similar plots for some other channels are presented in the Supplementary Materials Section S3. The plots clearly indicate that our method, with all the considered hyperparameter configurations, yielded significantly denoised estimates.

\vspace{-2pt}

\section{Musical Sound~Study} \label{sec:audio}
To further demonstrate the utility of the proposed method, we consider analyzing musical sounds that often exhibit sudden frequency changes over short periods, resulting in highly dispersed signals. For this study, we focus on a vocal music recording from 1934, originally published on a 78 RPM record of Hindustani classical music (one of India's two classical music traditions) and performed by eminent Ustad Amir Khan. A noisy copy of this recording was sourced from YouTube (link provided in the Supplementary Materials Section S1). From the recording, we extracted a 15-second segment representing a wide frequency range and saved it as a 16-bit WAV audio file, which represents noisy data. 

For our analysis, we divided each of the two audio file channels (left and right) into sections of 4096 data points to enhance computational efficiency. Each section was independently analyzed using the \emph{mixture-genlogit-doubleexp} and \emph{mixture-hypsec-doubleexp} methods (see Section~\ref{sec:simulation}) employing the Daubechies coiflets wavelet transform with five vanishing moments and periodic boundary conditions. The posterior estimates of the individual sections were then combined to reconstruct the posterior estimate of the entire audio~segment.

\begin{figure}[htbp!]
\centering
\includegraphics[width=\textwidth]{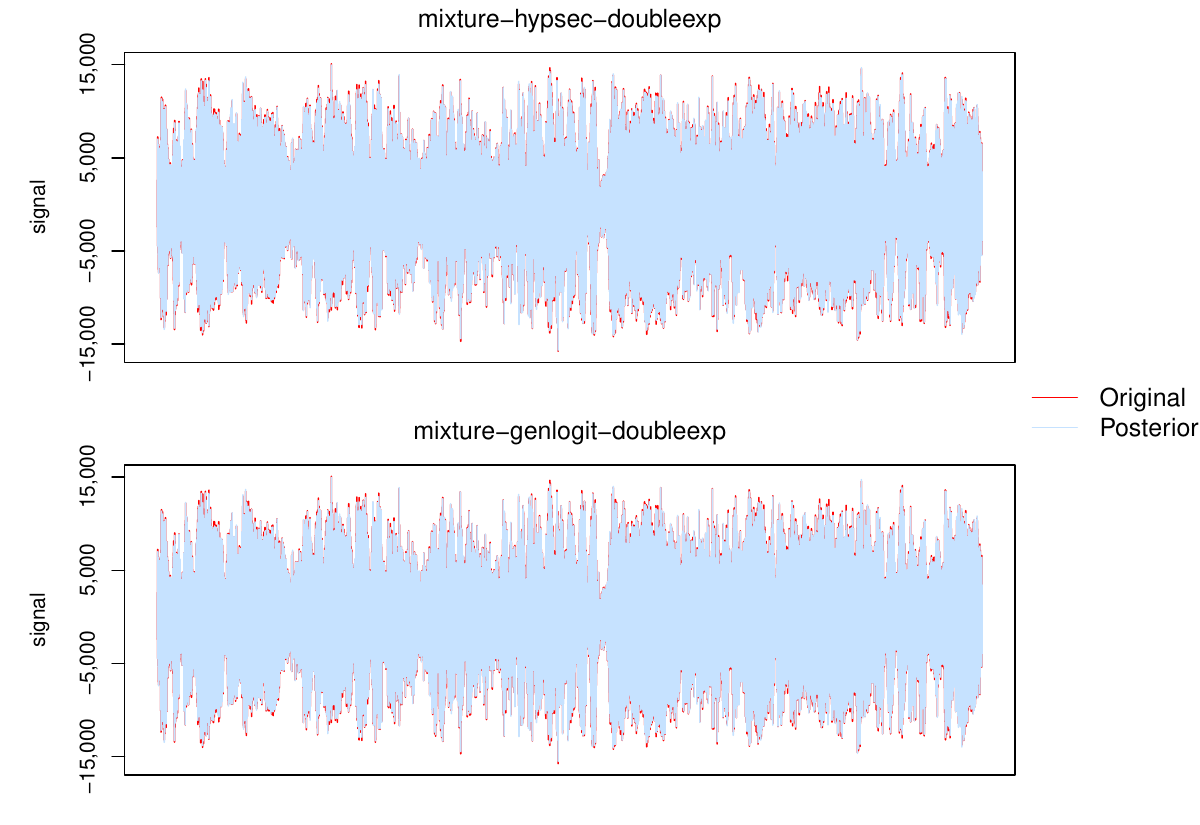}
\caption{Plots of the posterior means (in slate gray) of the right-channel signal of the chosen audio segment of the vocal music recording, superimposed on the noisy data (in red).}
\label{fig:song_post}
\end{figure}

In Figure~\ref{fig:song_post}, we present the posterior mean of the right-channel signal of the selected audio segment (in slate gray), superimposed on the noisy data (in black). The audio files corresponding to these posterior estimates, along with the original data and posterior estimates using the hard thresholding rule, are provided in the Supplementary Materials for auditory comparison and assessment. The posterior estimates show significant denoising and precise recovery of the signal. There is a slight systematic noise present in the posterior audio files, which is due to analyzing the whole segment in disjoint sections for computational convenience. That can be mitigated by analyzing larger sections or analyzing partially overlapping sections with weighted averaging (e.g., cross-fading or Hann windowing) to smoothly combine overlapping regions and reduce boundary artifacts.

\section{Discussion} \label{sec:discussion}
In this work, we proposed the use of nonlocal prior mixtures for wavelet-based nonparametric function estimation. The main innovations of our methodology are as follows:
\begin{enumerate}[label=(\alph*)]
\item We introduce a three-component spike-and-slab prior for the wavelet coefficients. This structure is particularly suited for modeling highly dispersed signals. The slab component is a mixture of two nonlocal priors---the MOM and IMOM priors, which offer enhanced adaptability to signal characteristics.
\item We propose flexible and previously unexplored hyperparameter specifications. These include generalized logit (or Richards), hyperbolic secant, and generalized normal decay specifications for the mixture probabilities, as well as a double exponential decay structure for the scale parameter. These enhancements provide improved flexibility and accuracy in modeling complex signal patterns, as demonstrated in our simulation study.
\item We implement our methodology within the R programming language \citep{R} as a package named NLPwavelet \citep{NLPwavelet}, which performs nonlocal prior (NLP)-based wavelet analysis.
\end{enumerate}

In the simulation study, using more dispersed versions of the Donoho--Johnstone test functions and various linear combinations of them, we compared the performance of the proposed approach with the existing two-component spike-and-slab mixture prior and demonstrated the superior flexibility of the proposed approach. Further, analysis using several novel hyperparameter configurations provided valuable insights into the relative advantages. Although no formal optimality results are available for these specific forms, their flexibility is supported by established theoretical principles for scale-dependent shrinkage, and their empirical performance, demonstrated in the simulation study of Section~4, provides strong evidence of their practical relevance.

Note that, in our empirical Bayes implementation, we did not encounter convergence failures. However, in smaller samples or high-noise scenarios, certain decay specifications—particularly those involving generalized normal parameters—showed greater variability in estimated hyperparameters. To reduce potential overfitting and identifiability issues, we recommend sensitivity checks across multiple decay specifications and inspection of hyperparameter estimates for plausibility. Such practices can help ensure the robustness of conclusions in applied settings.

A necessary limitation of the proposed approach is its higher computational cost relative to two-component mixture priors. Supplementary Materials Section S2 reports the average runtime (in seconds) for 24 analysis methods across the sample sizes $n$ considered in our simulation study. Within each method, the specification of the mixture probability has only a modest effect on runtime, while for the scaling parameter, \emph{doubleexp} options are generally slightly slower than \emph{polynom} options. The runtime growth of the mixture methods suggests approximately linear scaling with sample size. For instance, for~\emph{mixture-logit-polynom}, the~runtime for $n=1024$ is about 1.96 times that for $n=512$; for $n=2048$, it is about 1.96 times that for $n=1024$; and for $n=4096$, about 1.97 times that for $n=1024$. That near doubling of runtime when $n$ doubles is consistent across the other mixture methods too. This near doubling with each doubling of $n$ is consistent across other mixture methods, indicating a computational complexity of roughly $\mathcal{O}(n)$, which is generally considered good and efficient for large-scale problems. Nonetheless, the larger constant factor of the proposed three-component methods, relative to two-component methods, results in longer absolute runtimes, which is an anticipated trade-off for their enhanced modeling flexibility. While our simulations focus on one-dimensional signals, the linear scaling behavior suggests the approach can extend to higher-dimensional settings (e.g., 2D/3D images), subject to the corresponding increase in the number of coefficients.
 
Further, while the EEG and audio denoising examples illustrate the practical use of the proposed method, their evaluation is primarily qualitative due to the lack of ground truth reference signals. For EEG data, objective measures such as SNR improvement or reconstruction error cannot be computed reliably without a known clean signal. Similarly, in the audio example, perceptual quality metrics such as PESQ \citep{PESQ} or STOI \citep{taal_et_al_2010} require access to a clean reference, which was not available for the real-world recording used here. Future work, incorporating controlled experiments in which synthetic noise is added to high-quality EEG or audio recordings, would allow the computation of such quantitative metrics and hence direct, reproducible comparisons with existing denoising techniques, complementing the qualitative assessments presented in this study.

An obvious extension of the proposed approach is to adapt it to multidimensional wavelet regression, such as 2D or 3D image processing tasks.  Another possible avenue is to develop fully Bayesian hierarchical models by specifying prior distributions for the hyperparameters and employing Markov chain Monte Carlo tools or variational methods for posterior inference. In addition, one can explore combining wavelet decompositions with local polynomial regression to mitigate the boundary bias issues. Further, adapting our methodology to non-Gaussian or skewed data can be an interesting enterprise. We leave all these to future research.

\

\noindent \tb{Acknowledgements:} The authors thank the JAKAR High-Performance Cluster at the University of Texas at El Paso for providing computational resources free of charge.

\

\noindent \tb{Abbreviations:} The following abbreviations are used in this manuscript:
\\

\noindent 
\begin{tabular}{@{}ll}
MOM prior & Moment prior\\
IMOM prior & Inverse moment prior\\
NLP & Nonlocal prior\\
SNR & Signal-to-noise ratio\\
MSE & Mean squared error\\
EEG & Electroencephalogram
\end{tabular}


\section*{Appendix A}\label{app}

\subsection*{Appendix A.1. Proof of Result 1}
From the wavelet coefficient model (2) and the mixture prior of the wavelet coefficients in (3), using the Laplace approximation for the IMOM prior component, we get the marginal distribution of the empirical wavelet coefficients, $\widehat{d}_{lj}$, as~
\vspace{-12pt}
\begin{align*}
&\pi(\widehat{d}_{lj} | \sigma^{2},\bftheta,r,\nu) \\
&= \int \pi(\widehat{d}_{lj} | d_{lj},\sigma^{2},\bftheta,r,\nu) \pi(d_{lj} | \sigma^{2},\bftheta,r,\nu) dd_{lj} \\
&= \int (2\pi\sigma^{2})^{-\frac{1}{2}} \exp\left\{ -\frac{(\widehat{d}_{lj} - d_{lj})^{2}}{2\sigma^{2}} \right\}  \;  \left[ \gamma_{l}^{(1)} \widetilde{M}_{r} \left(\tau_{l}^{(1)}\sigma^{2}\right)^{-r-\frac{1}{2}} d_{lj}^{2r} \exp\left(-\frac{d_{lj}^{2}}{2\tau_{l}^{(1)}\sigma^{2}}\right) \right.\\
&\qquad \left. + \left(1 - \gamma_{l}^{(1)}\right)\gamma_{l}^{(2)} \frac{\left(\tau_{l}^{(2)}\sigma^{2}\right)^{\nu \over 2}}{\Gamma(\nu/2)} |d_{lj}|^{-\nu-1} \exp\left( -\frac{\tau_{l}^{(2)}\sigma^{2}}{d_{lj}^{2}} \right) + \left(1 - \gamma_{l}^{(1)}\right)\left(1 - \gamma_{l}^{(2)}\right) \delta(d_{lj}) \right] dd_{lj} \\
&=(2\pi\sigma^{2})^{-\frac{1}{2}} \left[ \gamma_{l}^{(1)} \widetilde{M}_{r} \left(\tau_{l}^{(1)}\sigma^{2}\right)^{-r-\frac{1}{2}} \int d_{lj}^{2r} \exp \left\{ -\frac{1}{2\sigma^{2}} ( (\widehat{d}_{lj} - d_{lj})^{2} + \frac{d_{lj}^{2}}{\tau_{l}^{(1)}})  \right\} dd_{lj} \right. \\
&\quad\left. + \left(1 - \gamma_{l}^{(1)}\right)\gamma_{l}^{(2)} \frac{\left(\tau_{l}^{(2)}\sigma^{2}\right)^{\nu \over 2}}{\Gamma(\nu/2)} \int |d_{lj}|^{-\nu-1} \exp\left\{ -\frac{(\widehat{d}_{lj} - d_{lj})^{2}}{2\sigma^{2}} - \frac{\tau_{l}^{(2)}\sigma^{2}}{d_{lj}^{2}} \right\} dd_{lj} \right.\\
&\left. \quad + \left(1 - \gamma_{l}^{(1)}\right)\left(1 - \gamma_{l}^{(2)}\right) \exp\left( -\frac{\widehat{d}_{lj}^{2}}{2\sigma^{2}} \right) \right]\\
&= (2\pi\sigma^{2})^{-\frac{1}{2}} \left[ \gamma_{l}^{(1)} \widetilde{M}_{r} \left(\tau_{l}^{(1)}\sigma^{2}\right)^{-r-\frac{1}{2}} \exp\left\{ -\frac{\widehat{d}_{lj}^{2}}{2\sigma^{2}\left(1+\tau_{l}^{(1)}\right)} \right\} \left( 2\pi\sigma^{2} \frac{\tau_{l}^{(1)}}{1+\tau_{l}^{(1)}} \right)^{1/2} \left( \sigma^{2} \frac{\tau_{l}^{(1)}}{1+\tau_{l}^{(1)}} \right)^{r}  \right. \\
&\quad\left. \sum_{i=0}^{r} \frac{(2r)!}{(2i)!(r-i)!2^{r-1}} \left(\frac{\frac{\tau_{l}^{(1)}}{1+\tau_{l}^{(1)}} \widehat{d}_{lj}}{\sqrt{\frac{\tau_{l}^{(1)}}{1+\tau_{l}^{(1)}} \sigma}}\right)^{2i} + \left(1 - \gamma_{l}^{(1)}\right)\gamma_{l}^{(2)} \frac{\left(\tau_{l}^{(2)}\sigma^{2}\right)^{\nu \over 2}}{\Gamma(\nu/2)} \exp\left( -\frac{\widehat{d}_{lj}^{2}}{2\sigma^{2}} \right) \sqrt{2\pi} \sigma_{*} h(d_{lj}^{*}(\widehat{d}_{lj})) \right. \\
&\quad\; \left. + \left(1 - \gamma_{l}^{(1)}\right)\left(1 - \gamma_{l}^{(2)}\right) \exp\left( -\frac{\widehat{d}_{lj}^{2}}{2\sigma^{2}} \right) \right] \qquad \text{(using Laplace approximation)} \\
&= \gamma_{l}^{(1)} \left(1+\tau_{l}^{(1)}\right)^{-r} M_{r}^{*}\left(\widehat{d}_{lj},\tau_{l}^{(1)},\sigma^{2}\right) \; \phi\left(\widehat{d}_{lj};0,\sigma^{2}\left(1+\tau_{l}^{(1)}\right)\right) + \left(1 - \gamma_{l}^{(1)}\right)\gamma_{l}^{(2)}  \frac{\left(\tau_{l}^{(2)}\sigma^{2}\right)^{\nu \over 2}}{\Gamma(\nu/2)} \\
&\qquad \phi\left(\widehat{d}_{lj};0,\sigma^{2}\right) \; \sqrt{2\pi} \sigma_{*} h(d_{lj}^{*}(\widehat{d}_{lj})) + \left(1 - \gamma_{l}^{(1)}\right)\left(1 - \gamma_{l}^{(2)}\right) \phi\left(\widehat{d}_{lj};0,\sigma^{2}\right).
\end{align*}
where
$$
M_r^{\star} \left(\widehat{d}_{lj}, \tau_l, \sigma^2\right) = \frac{1}{(2r-1)!!} \sum_{i=0}^r \frac{(2r)!}{(2i)!(r-i)!2^{r-i}} \left( \sqrt{\frac{\tau^{(1)}_l}{1+\tau^{(1)}_l}}\frac{\widehat{d}_{lj}}{\sigma} \right)^{2i},
$$
$$
h(d_{lj}) = |d_{lj}|^{-(\nu+1)} \exp\left\{ -\frac{1}{2\sigma^2} \left(d_{lj}^2 - 2d_{lj}\widehat{d}_{lj}\right) - \frac{\tau_l^{(1)}\sigma^2}{d_{lj}^2} \right\},
$$
$d_{lj}^{*}(\widehat{d}_{lj})$ is the global maxima of $h(d_{lj})$, and~$\sigma_*^2=-1/L_h''(d_{lj}^{*}(\widehat{d}_{lj}))$, with~$L_h(d_{lj})=\log(h(d_{lj}))$.

\subsection*{Appendix A.2. Proof of Result 2}
\vspace{-12pt}
\begin{align*}
\pi\left(d_{lj} | \sigma^{2},\bftheta,r,\nu,\bfy\right) &\propto \pi\left(\bfy | d_{lj},\sigma^{2},\bftheta,r,\nu\right) \; \pi\left(d_{lj} | \sigma^{2},\bftheta,r,\nu\right).
\end{align*}

The proportionality constant is
\begin{align*}
C &= \int \pi\left(\bfy | d_{lj},\sigma^{2},\bftheta,r,\nu\right) \pi\left(d_{lj} | \sigma^{2},\bftheta,r,\nu\right) dd_{lj} \\
&= (2\pi\sigma^{2})^{-\frac{1}{2}} \left[ \gamma_{l}^{(1)} \left(1+\tau_{l}^{(1)}\right)^{-r-{1\over 2}} M_{r}^{*}\left(\widehat{d}_{lj},\tau_{l}^{(1)},\sigma^{2}\right) \exp\left\{ -\frac{\widehat{d}_{lj}^{2}}{2\sigma^{2}\left(1+\tau_{l}^{(1)}\right)} \right\} \right.\\
&\left. \quad\;  + \left(1 - \gamma_{l}^{(1)}\right)\gamma_{l}^{(2)}  \frac{\left(\tau_{l}^{(2)}\sigma^{2}\right)^{\nu \over 2}}{\Gamma(\nu/2)} \exp\left( -\frac{\widehat{d}_{lj}^{2}}{2\sigma^{2}} \right) \sqrt{2\pi} \sigma_{*} h(d_{lj}^{*}(\widehat{d}_{lj})) \right. \\
&\quad\; \left. + \left(1 - \gamma_{l}^{(1)}\right)\left(1 - \gamma_{l}^{(2)}\right) \exp\left( -\frac{\widehat{d}_{lj}^{2}}{2\sigma^{2}} \right) \right]\\
&=(2\pi\sigma^{2})^{-\frac{1}{2}} \left(1 - \gamma_{l}^{(1)}\right)\left(1 - \gamma_{l}^{(2)}\right) \exp\left( -\frac{\widehat{d}_{lj}^{2}}{2\sigma^{2}} \right) \left[O_{lj}^{(1)} + O_{lj}^{(2)} + 1\right],
\end{align*}

\noindent Thus,
\allowdisplaybreaks
\begin{align*} 
&\pi\left(d_{lj} | \sigma^{2},\bftheta,r,\nu,\bfy\right) \\
=& \; {1\over C} \; \pi\left(\bfy | d_{lj},\sigma^{2},\bftheta,r,\nu\right) \; \pi\left(d_{lj} | \sigma^{2},\bftheta,r,\nu\right) \\
=&\; \frac{1}{(2\pi\sigma^{2})^{-\frac{1}{2}}  \left(1-\gamma_{l}^{(1)}\right)\left(1-\gamma_{l}^{(2)}\right)  \exp\left( -\frac{\widehat{d}_{lj}^{2}}{2\sigma^{2}}\right)  \left[O_{lj}^{(1)} + O_{lj}^{(2)} + 1\right]} (2\pi\sigma^{2})^{-\frac{1}{2}} \\
& \left[  \gamma_{l}^{(1)} \widetilde{M}_{r} \left(\tau_{l}^{(1)}\sigma^{2}\right)^{-r-\frac{1}{2}} \exp\left\{ -\frac{\widehat{d}_{lj}^{2}}{2\sigma^{2}\left(1+\tau_{l}^{(1)}\right)}\right\} d_{lj}^{2r} \exp\left\{ -\frac{1}{2\sigma^{2}\frac{\tau_{l}^{(1)}}{1+\tau_{l}^{(1)}}} \left( d_{lj} - \frac{\tau_{l}^{(1)}}{1+\tau_{l}^{(1)}} \widehat{d}_{lj} \right)^{2}\right\} \right. \\
& \left. +  \left(1-\gamma_{l}^{(1)}\right)\gamma_{l}^{(2)} \frac{\left(\tau_{l}^{(2)}\sigma^{2}\right)^{\nu \over 2}}{\Gamma(\nu/2)} \exp\left( -\frac{\widehat{d}_{lj}^{2}}{2\sigma^{2}} \right) \sqrt{2\pi} \sigma_{*} h(d_{lj}^{*}(\widehat{d}_{lj})) \phi\left(d_{lj} | d_{lj}^{*}(\widehat{d}_{lj}), \sigma_{*}^{2}\right) \right. \\
&\left. \qquad \exp\left\{\frac{\widehat{d}_{lj}d_{lj}}{\sigma^2} - \frac{d_{lj}^2}{2\sigma^2}\right\}\delta(d_{lj})  \right]  \\
=& \;\frac{O_{lj}^{(1)}}{O_{lj}^{(1)} + O_{lj}^{(2)} + 1} \; \frac{\widetilde{M}_r}{M_r^*\left(\widehat{d}_{lj}, \tau_l^{(1)}, \sigma^2\right)}  \; d_{lj}^{2r}  \; \exp\left\{ -\frac{1}{2\sigma^{2}\frac{\tau_{l}^{(1)}}{\left(1+\tau_{l}^{(1)}\right)}} \left( d_{lj} - \frac{\tau_{l}^{(1)}}{1+\tau_{l}^{(1)}} \widehat{d}_{lj} \right)^{2}\right\} \\
& + \frac{O_{lj}^{(2)}}{O_{lj}^{(1)} + O_{lj}^{(2)} + 1}  \phi\left(d_{lj} | d_{lj}^{*}(\widehat{d}_{lj}), \sigma_{*}^{2}\right)  +  \frac{1}{O_{lj}^{(1)} + O_{lj}^{(2)} + 1}   \exp\left\{\frac{\widehat{d}_{lj}d_{lj}}{\sigma^2} - \frac{d_{lj}^2}{2\sigma^2}\right\}\delta(d_{lj}) \\
=& \; p_{lj}^{(1)} \; \frac{\widetilde{M}_r}{M_r^*\left(\widehat{d}_{lj}, \tau_l^{(1)}, \sigma^2\right)}  \; d_{lj}^{2r}  \; \exp\left\{ -\frac{1}{2\sigma^{2}\frac{\tau_{l}^{(1)}}{\left(1+\tau_{l}^{(1)}\right)}} \left( d_{lj} - \frac{\tau_{l}^{(1)}}{1+\tau_{l}^{(1)}} \widehat{d}_{lj} \right)^{2}\right\}  + p_{lj}^{(2)}  \phi\left(d_{lj} | d_{lj}^{*}(\widehat{d}_{lj}), \sigma_{*}^{2}\right)   \\
& + \left(1-p_{lj}^{(1)}-p_{lj}^{(2)}\right)   \exp\left\{\frac{\widehat{d}_{lj}d_{lj}}{\sigma^2} - \frac{d_{lj}^2}{2\sigma^2}\right\}\delta(d_{lj}),
\end{align*}
from which Result 2 follows. 

\bibliographystyle{jasa}
\bibliography{mybib}

\end{document}